*Article*

# Two-Qubit Bloch Sphere

**Chu-Ryang Wie**

Department of Electrical Engineering, University at Buffalo, State University of New York, 230B Davis Hall, Buffalo, NY 14260, USA; wie@buffalo.edu

**Abstract:** Three unit spheres were used to represent the two-qubit pure states. The three spheres are named the base sphere, entanglement sphere, and fiber sphere. The base sphere and entanglement sphere represent the reduced density matrix of the base qubit and the non-local entanglement measure, concurrence, while the fiber sphere represents the fiber qubit via a simple rotation under a local single-qubit unitary operation; however, in an entangled bipartite state, the fiber sphere has no information on the reduced density matrix of the fiber qubit. When the bipartite state becomes separable, the base and fiber spheres seamlessly become the single-qubit Bloch spheres of each qubit. Since either qubit can be chosen as the base qubit, two alternative sets of these three spheres are available, where each set fully represents the bipartite pure state, and each set has information of the reduced density matrix of its base qubit. Comparing this model to the two Bloch balls representing the reduced density matrices of the two qubits, each Bloch ball corresponds to two unit spheres in our model, namely, the base and entanglement spheres. The concurrence–coherence complementarity is explicitly shown on the entanglement sphere via a single angle.

**Keywords:** two-qubit; Bloch sphere; quantum circuit; rotation operator; entanglement; coherence

**1. Introduction**

The Bloch sphere is a geometrical representation of pure single-qubit states as a point on the unit sphere [1]. Operations on single qubits commonly used in quantum information processing can be represented on the Bloch sphere. The north pole and the south pole of the Bloch sphere are defined as the orthonormal computational basis states |0⟩ and |1⟩, respectively, and an arbitrary single-qubit pure state, up to the global phase, is represented by a point on the unit sphere, thereby relating the superposition of the basis states to the angular coordinates of the point. Any unitary operation, taking an initial state to the final state of the single-qubit, is equivalent to a composition of one or more simple rotations on the Bloch sphere [1]. This Bloch sphere picture is elegant and powerful for the single qubit. It helps one visualize the superposition of quantum states in terms of the angular coordinates and the unitary operations on the state as rotations on the unit sphere. A similarly simple or widely accepted Bloch sphere model for two or more qubits is not available, although many efforts have been reported in the literature [2–7]. For two-qubit pure states, for example, a practical Bloch sphere model may try to depict the states using a point on the unit spheres where the linear superposition of basis states and the entanglement may be perceivable from the angle coordinates of the points, represent the local single-qubit gates as rotation operators on the local Bloch spheres, parameterize the entanglement property [8] using a small number of angle coordinates, and represent the joint two-qubit gates as some kind of controlled rotations on the Bloch spheres. Since the general multi-qubit unitary gates can ultimately be decomposed into various single qubit-gates and a two-qubit gate [1], a two-qubit Bloch sphere model can serve as a useful tool for the quantum circuit model of quantum computation.

The parameterization of two-qubit states requires seven parameters for pure states and fifteen parameters for general mixed states. The two-qubit state space can be described by a unit 7-sphere $S^7$ for pure states and by a four-dimensional special unitary group SU(4) for the mixed state density matrix and two-qubit operators. There have been various attempts to parameterize the two-qubit state space, including an explicit parameterization of SU(4) [2] and the construction of a special unitary group on two-qubit Hilbert space using the geometric algebra of a six-dimensional (6D) real Euclidean vector space [3]. Havel and Doran [3] pointed out that the even subalgebra of the geometric algebra for the three-dimensional Euclidean space of a single qubit is the quaternions [3]. They also developed the geometric algebra for a two-qubit system. Using a combinatorial graph consisting of six nodes (representing the three basis vectors of each qubit) and the bivectors as the edges connecting the nodes, they presented a complete picture for the two-qubit states and operations in the 6-D real Euclidean space. Here, the three nodes of each qubit and the edges (i.e., the bivectors) connecting the three nodes formed a local subgraph representing the local qubit states and operations [3]. In particular, Havel and Doran concluded that any two-qubit operations can be decomposed by what they called the "Cartan factorization," which says that any rotation in the 6D space can be uniquely written as a pair of rotations in the complementary 3D subspaces, followed by rotations in the three mutually orthogonal planes (i.e., bivectors or edges), each intersecting both 3-D subspaces (the so-called Cartan subalgebra), and finally followed by another pair of rotations on the 3-D subspaces [3]. The entanglement operation was represented by the middle part, i.e., by the Cartan subalgebra, and the entanglement was represented by a single angle parameter [3]. Mosseri and Dandoloff [4] applied Hopf fibration [9,10] and attempted to develop a Bloch sphere model for the two-qubit pure states. They proposed a three-dimensional ball with equal-concurrence concentric spherical shells as a possible model. Wie [5] extended this approach of Hopf fibration of the unit 7-sphere state space of two-qubit pure states and produced a complete parameterization. In the Hopf fibration [9,10], the two-qubit state space, i.e., the 7-sphere $S^7$, is mapped to a 4-sphere base space and a 3-sphere fiber space. Here, one qubit (called the base qubit) is assigned to the 4-sphere $S^4$ Hopf base space and the other qubit (called the fiber qubit) to the 3-sphere $S^3$ Hopf fiber space. Wang [6] applied the geometric algebra of Doran, Lasenby, and Gull [11] to analyze the space of two-qubit and three-qubit pure states, separating the local and non-local degrees of freedom. He used a single angle to represent the entanglement measure in terms of the von Neuman entropy, which at an angle of $\pi/2$ was 1 (i.e., maximal entanglement) [6]. The two-qubit mixed states were parameterized by expressing the 4 × 4 density matrix in terms of Dirac matrices, which included the pure states as a special case [7]. As expected, the geometric representation of the two-qubit states is much more complicated than the single-qubit case.

In this paper, an introduction of the two-qubit Bloch sphere model previously reported in Wie [5] is first given and a more complete version of the two-qubit pure state Bloch sphere model is presented, as well as an improved version of the entanglement sphere that includes an inner sphere of a radius pertinent to the value of concurrence, namely, the entanglement measure. This Bloch sphere model is based on the Hopf fibration, as reported in Wie [5], and can have two alternative versions because, in mapping the two-qubit state space $S^7$ using a Hopf fibration, the assignment of the two physical qubits to the Hopf base and fiber spaces is arbitrary. Hence, one can consider from which physical qubit's point of view to describe the Bloch sphere model for the two-qubit pure state. This qubit is chosen as the base qubit and may be considered the "local" qubit, while the fiber qubit may be considered the "remote" qubit. Therefore, given a two-qubit pure state, two alternative sets of Bloch spheres are possible, with three unit spheres each. Various examples of Bloch spheres as the two-qubit state progressively evolves in a two-qubit quantum circuit are provided. Finally, possible applications and implications of the model are discussed and summarized.

## 2. Summary of a Two-Qubit Bloch Sphere Model

The two-qubit Bloch sphere model initially presented in Wie [5] is summarized here. The two-qubit state vector $|\Psi\rangle$ is given by:

$$|\Psi\rangle = \alpha|00\rangle + \beta|01\rangle + \gamma|10\rangle + \delta|11\rangle, \tag{1}$$

where |00>, |01>, |10> and |11> are the two-qubit computational basis states; the complex amplitudes $\alpha$, $\beta$, $\gamma$, and $\delta$ satisfy a normalization condition and can be represented by a point on the 7-sphere $S^7$, which is the unit sphere in Euclidean 8-space. This 7-sphere can be mapped to a 4-sphere $S^4$, the so-called Hopf base, and a fiber $S^3$ using the Hopf fibration process [9,10]. The 4-sphere $S^4$, the Hopf base,

is a unit sphere in the Euclidean 5-space given as $x_0^2 + x_1^2 + x_2^2 + x_3^2 + x_4^2 = 1$ in the Cartesian coordinates. It was assumed that one qubit is assigned to the $S^4$ base and the other qubit to the $S^3$ fiber. Here, the qubit assigned to the base is referred to as the base qubit and the other qubit is referred to as the fiber qubit.

A general single-qubit state can be written as the eigenspinor of a Pauli matrix for direction specified by $\theta$ and $\phi$ in spherical coordinates, which is given as a 2-vector in a complex field. This spinor represents a point on the $S^2$ Bloch sphere. If the complex imaginary unit "$i$" of the single-qubit spinor is replaced by a variable, pure-imaginary unit quaternion "$t$," where $t^2 = -1$, this spinor in the quaternion field can represent a point on the 4-sphere $S^4$ with four degrees of freedom (two from the angle coordinates and two from the variable pure-imaginary unit quaternion $t$). This spinor in the quaternion field is written as [5]:

$$|\tilde{\psi}\rangle = \begin{pmatrix} \cos\frac{\theta}{2} \\ \sin\frac{\theta}{2} e^{t\phi} \end{pmatrix}. \tag{2}$$

Here, the tilde on the *ket*-vector $|\tilde{\psi}\rangle$ is to indicate that Equation (2) is not a usual quantum state, which is commonly written as a spinor in a complex field. Note that Equation (2) is a normalized 2-vector. The general pure-imaginary unit quaternion $t$ can be parameterized with another two angles, $\chi$ and $\xi$:

$$t = i \sin\chi\cos\xi + j \sin\chi\sin\xi + k \cos\chi, \tag{3}$$

where $i$, $j$, and $k$ are the anti-commuting imaginary units of a quaternion. The spinor in a quaternion field (Equation (2)) can be represented by a point on the Hopf base 4-sphere and can be shown on two 2-spheres. One 2-sphere with angle coordinates ($\theta$, $\phi$) is named "the base sphere" and the other 2-sphere with angle coordinates ($\chi$, $\xi$) is named "the entanglement sphere." The entanglement sphere is limited to the northern hemisphere ($\chi \leq \pi/2$), which is the result of Equation (2) [5].

For the single-qubit pure state, the zero superposition states (classical bits or the computational basis states) and the uniform superposition (maximum superposition) states are shown on the Bloch sphere in Figure 1. For a two-qubit pure state, in order to depict the $S^4$ Hopf base space, the $S^4$ Cartesian coordinates, $x_0$, $x_1$, $x_2$, $x_3$, and $x_4$ in the Euclidean 5-space were defined in terms of four angles $\theta$, $\phi$, $\chi$ and $\xi$ in Wie [5]. The two unit-spheres, *i.e.*, the *base* sphere and the *entanglement* sphere, together were shown to represent the reduced density matrix of the *base* qubit and the entanglement measure *concurrence* of the bipartite state [5]. On these two spheres, the separable states and the maximally entangled states (MES) are shown in Figure 2.   It is interesting to note that the *entanglement* sphere in Figure 2 for the separable (*concurrence*=0) and MES (*concurrence*=1) states is similar in appearance to the single-qubit Bloch sphere in Figure 1 for the zero and uniform superposition states, respectively.

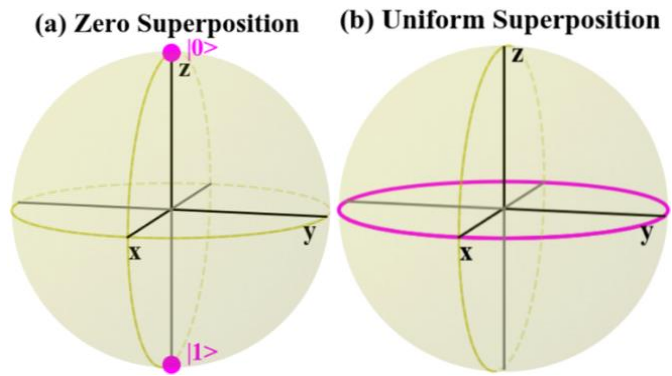

**Figure 1.** Single-qubit Bloch sphere: (**a**) the zero superposition states, |0⟩ and |1⟩, and (**b**) the uniform superposition states. The sine value of the polar angle of the Bloch vector is zero for (a), and unity for (b).

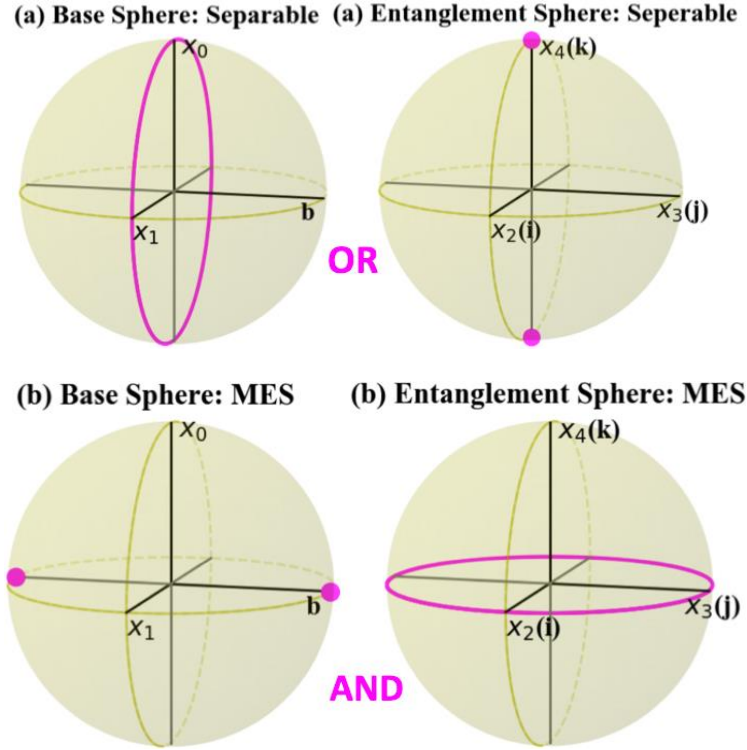

**Figure 2.** Two-qubit Bloch spheres, showing the base and entanglement spheres of the $S^4$ Hopf base for (**a**) the separable states and (**b**) the maximally entangled states (MES). For (a), the coordinates of either sphere means that the bipartite state is separable, and for (b), both sphere conditions must be satisfied for MES.

A point on the $S^3$ fiber space can be represented by a unit quaternion $q_f$, which, together with the spinor Equation (2) for Hopf base $S^4$, completes the Bloch sphere parameterization for a two-qubit pure state:

$$\begin{pmatrix} \cos\frac{\theta}{2} \\ \sin\frac{\theta}{2}e^{t\phi} \end{pmatrix} q_f \ = \ \begin{pmatrix} \alpha + \beta j \\ \gamma + \delta j \end{pmatrix} \text{ or } \begin{pmatrix} \alpha + \gamma j \\ \beta + \delta j \end{pmatrix}. \tag{4}$$

Here, the first 2-vector on the right-hand side is when qubit-A is assigned to the Hopf base, and the second 2-vector is when qubit-B is assigned to the Hopf base. It should be noted that the complex amplitudes of $\alpha$, $\beta$, $\gamma$, and $\delta$ must use the quaternion imaginary unit "$k$" as their complex imaginary unit to work correctly in the quaternion algebra here. That is, a complex number $\alpha$ with the real part $\alpha_r$ and imaginary part $\alpha_i$, complex($\alpha_r$, $\alpha_i$), must become quaternion($\alpha_r$, 0, 0, $\alpha_i$) to use the complex number $\alpha$ in the quaternionic Equation (4). Likewise for $\beta$, $\gamma$, and $\delta$. The unit quaternion $q_f$ for the $S^3$ fiber space is parameterized by three angles: ($\theta_f,\varphi_f$) for the fiber sphere and $\zeta_f$ for a phase factor:

$$q_f \ = \ \left(\cos\frac{\theta_f}{2} + \sin\frac{\theta_f}{2}e^{k\phi_f}j\right)e^{k\zeta_f}. \tag{5}$$

As either of the two qubits may be assigned to the Hopf base, the present model consists of two alternative sets of three unit spheres, with each set having one of the two qubits assigned to the base and the other qubit to the fiber. It should be noted that only the Hopf base $S^4$ (consisting of the base sphere and the entanglement sphere) has all local information about the base qubit and the non-local information of entanglement for the bipartite state.

The Cartesian coordinates of the base sphere are given in terms of the angular coordinates as ($x_1$, $b$, $x_0$) = (sin$\theta$cos$\varphi$, sin$\theta$sin$\varphi$, cos$\theta$). The entanglement sphere may be represented by $t$, $bt$, or both. In this paper, we modify the entanglement sphere relative to the one presented in Wie [5], shown in Figure 2, by including both $t$ (unit radius, the outer sphere) and $bt$ (radius $|b|$, the inner sphere). The inner entanglement sphere has the Cartesian coordinates $(x_2, x_3, x_4) = b(\sin\chi\cos\xi, \sin\chi\sin\xi, \cos\chi) = bt$. As will be discussed later, the entanglement measure concurrence is given by $\sqrt{(x_2^2 + x_3^2)} = |b|\sin\chi$, and the imaginary part of coherence of the base qubit is given by $x_4 = b\cos\chi$. The Bloch sphere coordinates are obtained from the four complex amplitudes of the bipartite states in Equation (1) as follows:

If qubit-A is in the Hopf base and qubit-B is in the fiber, the Bloch sphere coordinates are obtained from (qubit-A in the $S^4$ Hopf base):

$$1 + x_0 = 2(|\alpha|^2 + |\beta|^2), \tag{6}$$

$$x_1 + bt = 2(\bar{\alpha}\gamma + \bar{\beta}\delta). \tag{7}$$

The angular coordinates $(\theta_f, \varphi_f-2\zeta_f)$ of the fiber sphere (qubit-B) and the phase factor $\zeta_f$ are obtained by equating the definition of $q_f$ in Equation (5) to:

$$q_f = \cos\frac{\theta}{2}(\alpha + \beta j) + \sin\frac{\theta}{2}e^{-t\phi}(\gamma + \delta j). \tag{8}$$

If qubit-B is in the Hopf base and qubit-A is in the fiber, the Bloch sphere coordinates are obtained from (qubit-B in the $S^4$ Hopf base):

$$1 + x_0 = 2(|\alpha|^2 + |\gamma|^2), \tag{9}$$

$$x_1 + bt = 2(\bar{\alpha}\beta + \bar{\gamma}\delta), \tag{10}$$

$$q_f = \cos\frac{\theta}{2}(\alpha + \gamma j) + \sin\frac{\theta}{2}e^{-t\phi}(\beta + \delta j). \tag{11}$$

## 3. Improved Entanglement Sphere and All Three Bloch Spheres

In the initial model [5], the entanglement sphere had only the *t*-vector represented as a unit sphere. In this paper, the *bt*-vector is included and it is represented using an inner sphere of radius $|b|$ inside the *t* sphere of unit radius. Keeping the outer sphere of unit radius helps gauge the size of the inner sphere visually. It may also be useful for estimating the concurrence and the magnitude of the imaginary part $x_4$ of the coherence quickly and it allows for the *t* coordinates to be displayed. This way, all the related parameters for the concurrence and the imaginary part of the coherence are presented in one sphere without the need to refer to the base sphere for the *b*-value. The intercept of the *t*-vector with the inner sphere determines both the non-local concurrence as the radius $c$ of the circle parallel to the equatorial $x_2$-$x_3$ plane, $c = \sqrt{(x_2^2 + x_3^2)} = |b|\sin\chi$, and the imaginary part of the coherence of the base qubit up to the sign, $x_4 = b\cos\chi$. This improved version of the entanglement sphere is shown in Figure 3 and all three Bloch spheres are shown in Figure 4.

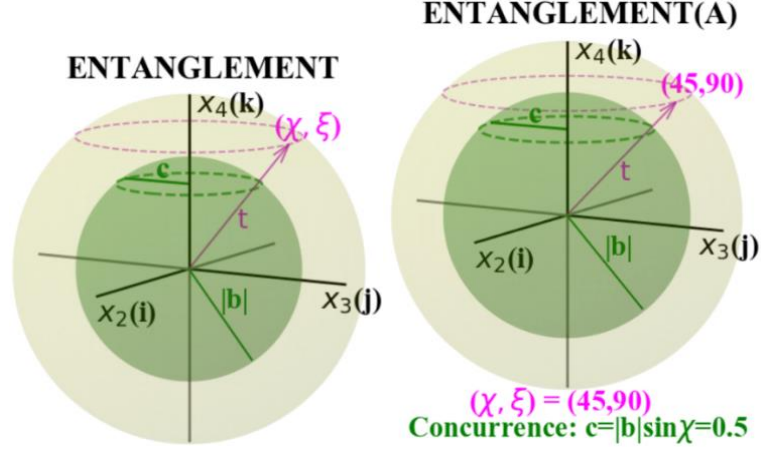

**Figure 3.** The improved entanglement sphere has two spheres: the outer sphere of unit radius (yellow) and the inner sphere of radius |b| (green). The pure-imaginary unit quaternion *t* with angular coordinates ($\chi,\xi$) meets the outer sphere and the dashed magenta horizontal circle is drawn. *t* intersects the inner sphere where the dashed green horizontal circle of radius *c*, concurrence, is drawn. The z-coordinate of the green dashed circle (on the inner sphere) is the $x_4$, up to the sign, of the reduced density matrix (Equation (12)). The intersection of *t* with the inner sphere gives the coordinate ($x_2$, $x_3$, $x_4$). The sphere on the right shows an example of an entanglement sphere indicating which qubit is the base qubit (qubit-A), with values displayed for the angular coordinates and the concurrence.

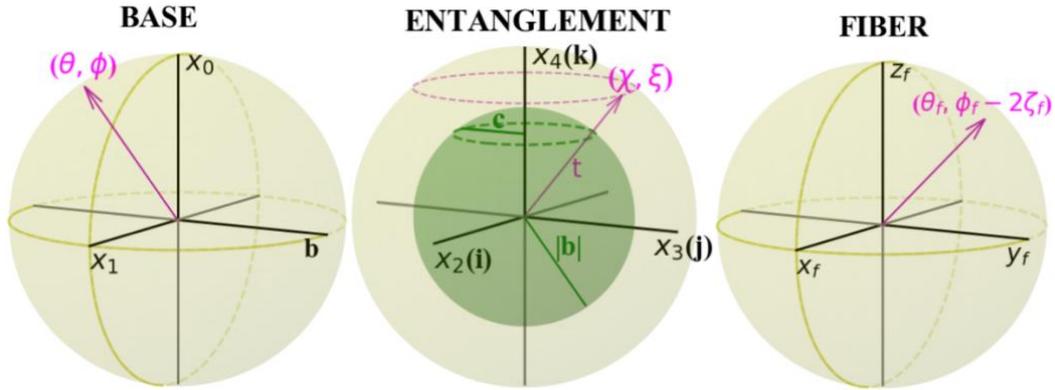

**Figure 4.** The two-qubit Bloch sphere: the base sphere with angle coordinates ($\theta,\phi$), the entanglement sphere ($\chi,\xi$), the fiber sphere ($\theta_f$, $\varphi_f - 2\zeta_f$), and the phase factor $\zeta_f$. The entanglement sphere illustrated is for concurrence $c = 0.5$ and $t = (j + k)/\sqrt{2}$.

The reduced density matrix of the base qubit is obtained by tracing out the fiber qubit:

$$\rho_{base} = Tr_{fiber}(|\Psi\rangle\langle\Psi|) = \frac{1}{2}\begin{pmatrix} 1 + x_0 & x_1 - x_4 k \\ x_1 + x_4 k & 1 - x_0 \end{pmatrix}, \quad (12)$$

where $x_4 = b\cos\chi$, and $|x_4| < |b|$ for a mixed state and $|x_4| = |b|$ for a pure state because the angle $\chi = 0$ is a separable bipartite state (see Figure 2). Equation (12) is obtained from Equations (1), (4), (6), and (7) when qubit-A is in the Hope base; or (1), (4), (9) and (10) when qubit-B is in the Hopf base. The base sphere depends on the measurement statistics $x_0$, according to the Born rule, on the real part of the coherence $x_1$ and on the intermediary parameter *b* for the imaginary part $x_4$ of the coherence and the concurrence *c* which are shown on the entanglement sphere.

The base sphere represents the base qubit, which for a separable state is precisely the single-qubit pure-state Bloch sphere, and for an entangled bipartite state, is the mixed state Bloch ball which is stretched along the y-axis (b-axis) into a unit sphere. The latter point can be seen from $\rho_{base} = \frac{1}{2}(I + \mathbf{r}\cdot\boldsymbol{\sigma}) = \frac{1}{2}(I + x_1\sigma_x + x_4\sigma_y + x_0\sigma_z)$, where *I* is the 2 by 2 unit matrix, $\mathbf{r} = (r_x, r_y, r_z) = (x_1, x_4, x_0)$ is the Bloch vector and $\boldsymbol{\sigma}$ is the Pauli matrix $\boldsymbol{\sigma} = (\sigma_x, \sigma_y, \sigma_z)$. That is, the base sphere has the y-axis expanded relative to the Bloch ball ($x_1$, $x_4$, $x_0$) such that it is a perfect unit sphere: $x_1^2 + b^2 + x_0^2 = 1$. Conversely, for an entangled bipartite state, a mixed-state base-qubit $\rho_{base}$ Bloch ball may be obtained by compressing the unit base sphere along the $y$-axis by the factor $\cos\chi$.

The inner entanglement sphere represents the imaginary part of coherence of the base qubit, $x_4 = b\cos\chi$ up to the sign and the concurrence $c = |b|\sin\chi$ of the bipartite state. These two parameters, one local to the base qubit and one non-local, can be seen related in the representation of the base qubit state by the spinor in the quaternion field (Equation (2)). Taking an outer product of Equation (2) gives the following:

$$|\tilde{\psi}\rangle\langle\tilde{\psi}| = \frac{1}{2}\begin{pmatrix} 1 + x_0 & x_1 - bt \\ x_1 + bt & 1 - x_0 \end{pmatrix} = \frac{1}{2}\begin{pmatrix} 1 + x_0 & x_1 - x_4 k - c e^{k\xi} i \\ x_1 + x_4 k + c e^{k\xi} i & 1 - x_0 \end{pmatrix}. \quad (13)$$

The $x_4$ and $c$ are the imaginary parts of the off-diagonal elements of Equation (13). As can be seen in Figure 3, the intersection of the $t$-vector with the inner entanglement sphere (the $bt$-sphere) gives both of these imaginary off-diagonal elements, including the azimuth angle $\xi$ which has no obvious physical interpretation. This relation is also suggested in the definition of concurrence [8]:

$$c = \sqrt{2(1 - Tr\rho_{base}^2)} = \sqrt{1 - x_0^2 - x_1^2 - x_4^2} = \sqrt{x_2^2 + x_3^2}. \quad (14)$$

Therefore, the $S^4$ Hopf base encodes the imaginary part of coherence $x_4$ up to the sign and the concurrence $c$, including the phase factor $\xi$ on the inner entanglement sphere. For a given $b$-value in the base sphere, a rising concurrence $c$ (by the increasing polar angle $\chi \in [0, \frac{\pi}{2}]$ on the entanglement sphere) means a falling magnitude $|x_4|$ of the imaginary part of the coherence, and a falling $c$ means a rising $|x_4|$. This illustrates the concurrence-coherence complementarity which is further discussed later in Sec.5.

The fiber sphere, along with the phase factor $\zeta_f$, represents the fiber qubit geometrically as a composition of simple rotations in response to a local single-qubit unitary operation. This will be shown in the next section by applying a single-qubit rotation gate on the fiber qubit. For a separable bipartite state, the fiber sphere is precisely the Bloch sphere of the fiber qubit. For an entangled bipartite state, the fiber sphere has no information on the reduced mixed-state density matrix of the fiber qubit and its main role seems to be to help the three Bloch spheres represent the bipartite state vector precisely, via Equation (4).

## 4. Bloch Sphere Examples with a Two-Qubit Quantum Circuit

In this section, some examples of the Bloch sphere coordinates of partially entangled two-qubit states are discussed in terms of the rotation angles of the unitary gates in a two-qubit quantum circuit. In each case, two sets of Bloch spheres, one set with qubit-A (the top qubit in the quantum circuit) assigned to the Hopf base and the other case with qubit-B (the bottom qubit in the quantum circuit) assigned to the Hopf base are presented.

Figure 5 shows the entangling circuit with $x$- or $y$-rotation by angle $\eta$, $R_{x,y}(\eta)$, producing a linear superposition of basis states in the control-qubit, qubit-A. Note that the degree of superposition of the $|0\rangle$ and $|1\rangle$ states is $\sin\eta$ after this gate. The controlled $x$- or $y$-rotation, C-$R_{x,y}(\omega)$, entangles the two qubits. The concurrence is given by $\sin\eta \sin(\omega/2)$ after these two gates.

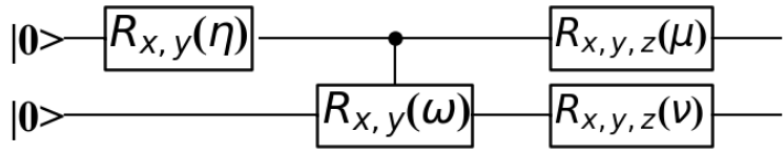

**Figure 5.** Two-qubit quantum circuit to generate partial entanglement between the two qubits with 0 ≤ concurrence ≤ 1. The gates are the rotation operators. Concurrence = 0 for separable states and 1 for maximally entangled states (MES). For this circuit, concurrence = $\sin(\eta)\sin(\omega/2)$ and $\sin\eta$ is the degree of initial superposition of the control qubit.

Figure 6 shows the two sets of Bloch spheres after a $R_x(60°)\otimes I$ gate to induce the superposition in the control qubit-A and a $C\text{-}R_y(70°)$ gate to entangle the two qubits. The rotation angles are given in degrees. First, note that as expected, the concurrence value is the same in the two sets (top row and bottom row) of Bloch spheres with a value equal to $\sin(60°)\sin(70°/2) \approx 0.5$. If the control qubit (qubit-A) is assigned to the base sphere, marked as BASE(A) in the figure, the base sphere shows that after the two gates, its rotation angle is precisely as if it is a free Bloch sphere applied only with the (initial) rotation of $R_x(60°)$. This is true for an initial $x$-rotation of the base sphere because the subsequent controlled rotation does not need to change its $b$-coordinate. However, if the initial rotation is a $y$-rotation, which will keep the $b$-coordinate at zero after the initial rotation, then the subsequent controlled rotation will need the $b$-coordinate to develop a non-zero value in the base sphere to have a non-zero concurrence. Hence, after an initial $y$-rotation, the subsequent controlled rotation will cause the base sphere to rotate as if it is coupled to the entanglement sphere (note: this example is not shown here, but one can verify it with the software provided [12]). Therefore, the initial $x$-rotation and $y$-rotation of the control qubit act differently due to a subsequent controlled-gate.

If the control qubit (qubit-A) is assigned to the fiber sphere, FIBER(A) in Figure 6, after the two gates, the fiber sphere rotates by a rotation angle that is smaller than the angle specified in the initial rotation operator $R_x(60°)$, although the sense and the axis of rotation are the same as the rotation operator. This rotation of the fiber sphere after the two gates occurs as if the fiber sphere is coupled to the base sphere. The BASE(B) coordinate is not a simple result of the $y$-rotation in the $C\text{-}R_y(70°)$ gate. The Bloch sphere coordinates in Figure 6 are used as the starting point in Figures 7 and 8 to discuss the effects of a subsequent single-qubit gate.

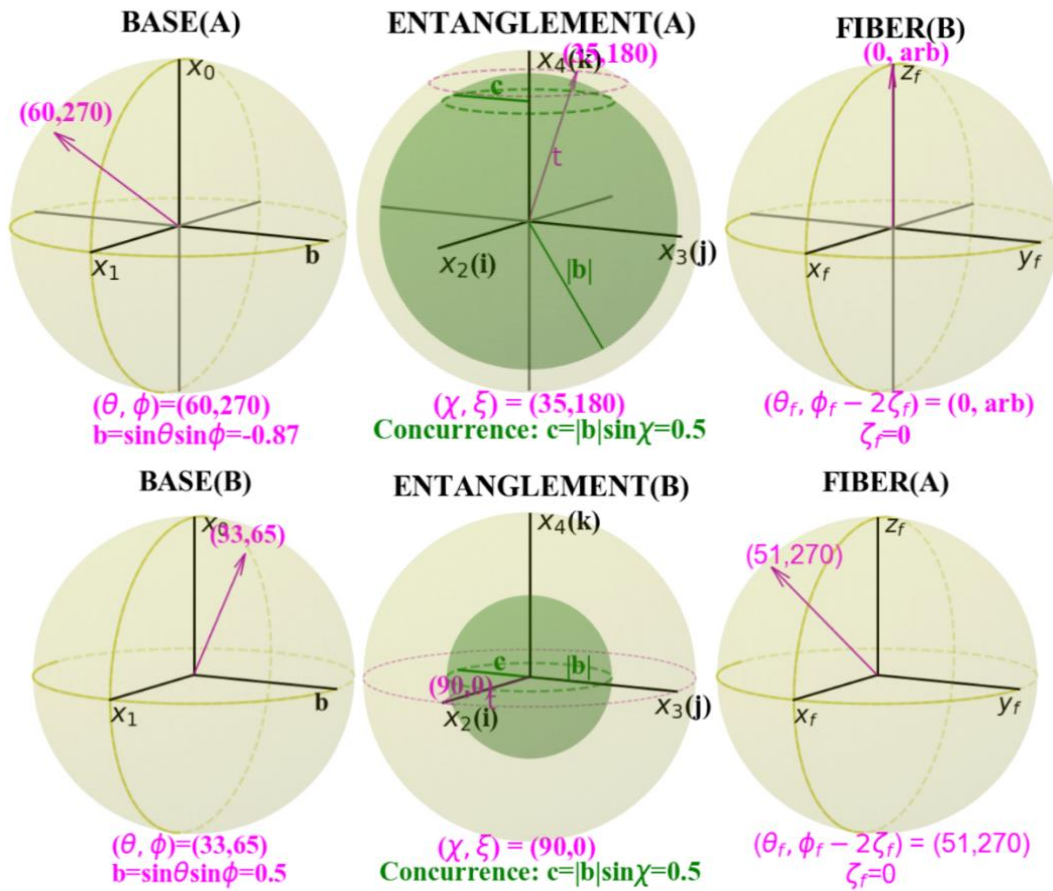

**Figure 6.** The two sets of Bloch spheres where qubit-A is the control qubit, after $R_x(60°)\otimes I$ → $C\text{-}R_y(70°)$ in the circuit of Figure 5, where the rotation angles are $\eta = 60°$, $\omega = 70°$, and $\mu = \nu = 0°$. Top row: qubit-A is the base qubit. Bottom row: qubit-B is the base qubit.

In the entangled bipartite state, the effect of a local unitary operation on qubit-A is shown in Figure 7. On the Bloch sphere coordinates given in Figure 6, a *y*-rotation by 90° is applied to qubit-A. First, since no local rotation can change the entanglement, this results in the same concurrence value in Figure 7 as in Figure 6. Furthermore, this *y*-rotation of qubit-A is seen as a simple *y*-rotation on the BASE(A) and FIBER(A) spheres, and the reduced density matrix of qubit-B is not affected, as shown by the BASE(B) and ENTANGLEMENT(B) spheres in Figures 6 and 7. Second, when the local *y*-rotation is applied to the fiber qubit, the rotation occurs only on the fiber sphere, as shown by FIBER(A) in Figure 7 relative to FIBER(A) in Figure 6. In other words, the base and entanglement sphere coordinates, BASE(B) and ENTANGLEMENT(B), are exactly the same in Figures 6 and 7, while FIBER(A) in Figure 7 is an exact 90° rotation about the *y*-axis from that in Figure 6. Third, if the local unitary operation is applied to the base qubit, BASE(A), and does not change its *y*-coordinate (the *b*-value), as is the case with a *y*-rotation, then the entanglement sphere remains fixed (ENTANGLEMENT(A) in Figures 6 and 7) and the base sphere rotates as if it is a free sphere (BASE(A) in Figure 7 relative to that in Figure 6). In this case, the fiber sphere adjusts itself (FIBER(B) in Figures 6 and 7) so that the three spheres correctly yield the bipartite state according to Equation (4). In this case, the fiber sphere (FIBER(B) in Figure 7) is just a mathematical placeholder with no apparent physical interpretation.

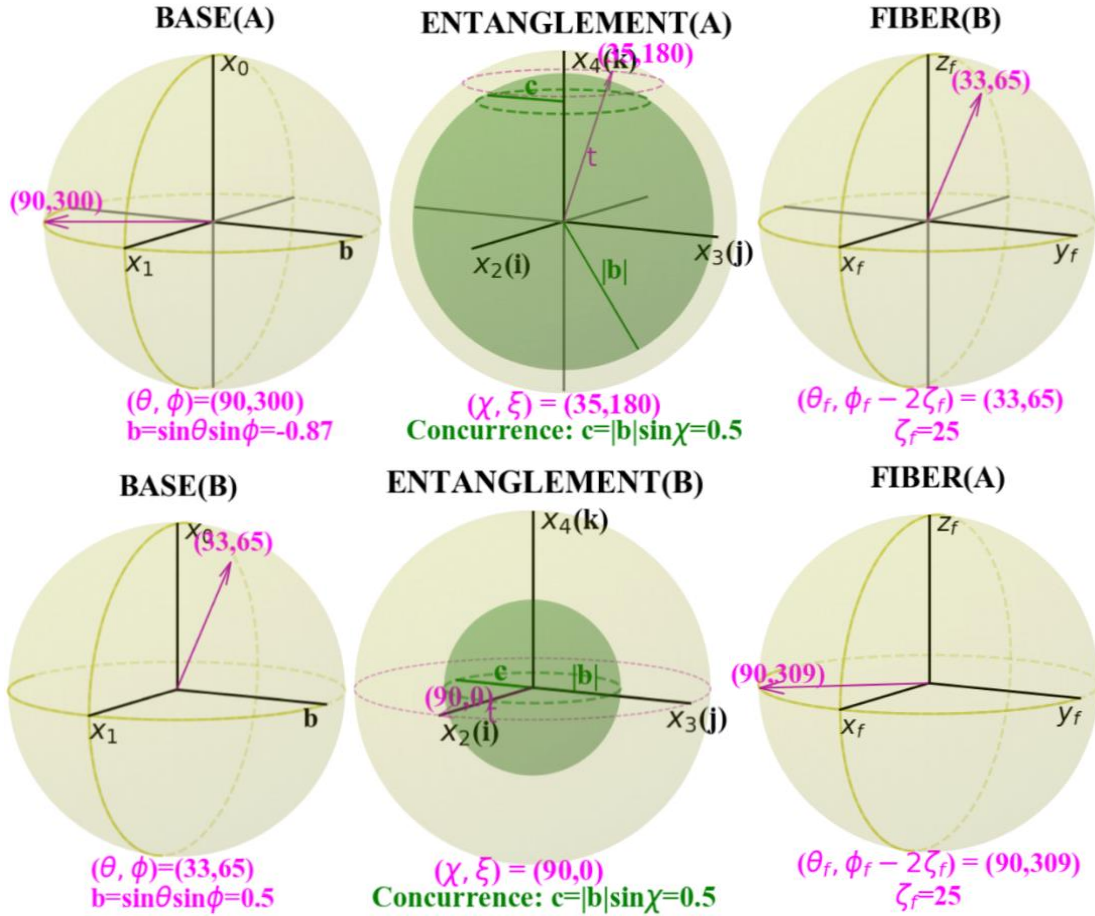

**Figure 7.** The two sets of Bloch spheres after the partial entanglement and a local rotation induced by $R_x(60°)\otimes I \rightarrow C\text{-}R_y(70°) \rightarrow R_y(90°)\otimes I$ in the circuit of Figure 5, where the rotation angles are $\eta = 60°$, $\omega = 70°$, $\mu = 90°$, and $\nu = 0°$. For the top row, qubit-A is the base qubit, and for the bottom row, qubit-B is the base qubit. After Figure 6, $R_y(90°)$ is applied to qubit-A.

Figure 8 shows examples of the effects of a *b*-changing rotation (*z*-rotation in the top row, and *x*-rotation in the bottom row) on the base qubit, applied to the entangled bipartite state of Figure 6 (top row). The top row of Figure 6 was compared with both rows of Figure 8. The *b*-changing *z*-rotation $R_z(90°)$ on qubit-A (top row of Figure 8) rotates the base sphere in the same sense about the *z*- axis (i.e., the right-hand rule) but by a smaller amount (55° in BASE(A) about the *z*-axis instead of 90° of $R_z(90°)$), and the ENTANGLEMENT(A) sphere adjusts to give *c* = 0.5, the same as in Figure 6, and yields $x_4$ = 0. The *b*-changing *x*-rotation $R_x(−90°)$ on qubit-A (bottom row of Figure 8) rotates the base sphere about the *x*-axis by −105° in BASE(A) rather than −90°, while ENTANGLEMENT(A) adjusts to *c* = 0.5 and $|x_4|$ = 0.5. FIBER(B) also adjusts. Hence, the *b*-changing rotations on the base sphere result in a complicated change. Therefore, the following identities may be useful for a *b*-changing *x*- or *z*-rotation to express it in terms of the *b*-preserving *y*-rotation when performing a local operation on the base qubit:

$$R_x(\mu) = R_z(-90°)R_y(\mu)R_z(90°); \; R_z(\mu) = R_x(90°)R_y(\mu)R_x(-90°).$$

The rotation angles are given in degrees.

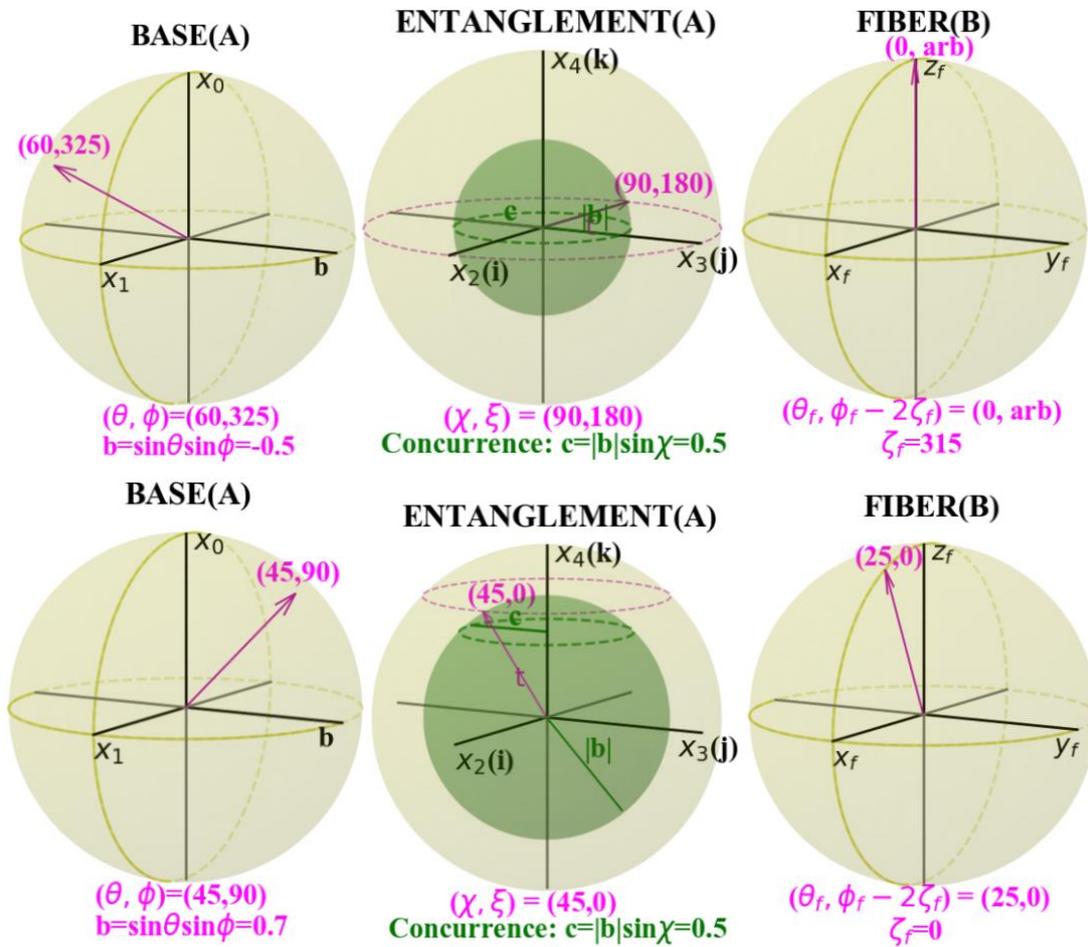

**Figure 8.** Base qubit local operation examples: $R_x(60°)\otimes I$ → $C$-$R_y(70°)$ → $R_z(90°)\otimes I$ (top row) and $R_x(60°)\otimes I$ → $C$-$R_y(70°)$ → $R_x(−90°)\otimes I$ (bottom row). After the top row of Figure 6, $R_z(90°)$ or $R_x(−90°)$ is applied to qubit-A. Compare the base and entanglement spheres with the corresponding ones in the top row of Figure 6. A local *z*- or *x*-rotation of the base sphere affects both the base and entanglement coordinates as the *b*-value is changed by the rotation.

## 5. Discussion

In the presented Bloch spheres, both the reduced density matrices, as well as the entanglement measure, concurrence, are represented consistently for the entire range of entanglement, from separable to the maximally entangled states. A reduced density matrix has three parameters (one diagonal element and the two off-diagonal elements, namely, the real and imaginary parts). The concurrence requires another parameter. However, the concurrence and the reduced density matrix are related by [8]:

$$c = \sqrt{2(1 - Tr\rho^2)},$$

where $\rho$ is the reduced density matrix, leaving three parameters for the Bloch spheres to encode for each qubit. A unit sphere has two degrees of freedom. In the presented model, the three parameters and the concurrence–coherence relation are represented on the two unit spheres, i.e., the base sphere and the entanglement sphere. To represent both reduced density matrices, two alternative sets of Bloch spheres are used, where each set is relative to the qubit assigned to the base sphere.

With the two alternative sets of Bloch spheres, it is convenient to consider one qubit a "local" qubit and assign it to the base sphere and consider the other qubit a "remote" qubit and assign it to the fiber sphere. For a separable bipartite state, the base and fiber spheres are precisely the single-qubit Bloch spheres of each qubit. For an entangled bipartite state, the base sphere, together with the entanglement sphere, can carry information on the local, base qubit (i.e., the reduced density matrix), as well as the entanglement information of the bipartite pure state, while the fiber sphere completes the parameterization via Equation (4). A single-qubit unitary operation on the base qubit may be studied by examining the base and entanglement spheres only ($x_0$ and $x_1$ on BASE(A) and $x_4$ and concurrence on ENTANGLEMENT(A) in Figure 7). The entanglement property, concurrence, shown on the entanglement sphere, remains unchanged by any local single-qubit operation. In an entangled state, the reduced density matrix is the only entity that is affected by a local unitary interaction on the base qubit, while the fiber sphere (FIBER(B) in Figure 7) has no information on the fiber qubit that can be measured. This is consistent with interpreting the fiber qubit to be a remote object. Furthermore, if the local unitary operation on the base qubit is such that its $y$-coordinate (i.e., the $b$-value) remains unchanged, as is the case in any $y$-rotation, the entanglement sphere, which encodes $x_4$ and concurrence, remains constant, leaving only the base sphere to rotate.

When measuring the spin direction of the local, base qubit, the probability is $\cos^2(\theta/2)$ for the spin-up |0⟩ state, whether the bipartite state is entangled or separable. A potentially useful application of our Bloch sphere model is the bipartite entanglement experiment on two interacting two-level atoms using double coherent ultrashort laser pulses, as reported by Yu and Li [13]. The entanglement evolution, the so-called entanglement sudden death and recurrence, may be investigated with the aid of the base sphere and the entanglement sphere. These two unit spheres can display, and may add insight to, the evolution of the reduced density matrices (for coherence), as well as the concurrence.

Next, the Bloch sphere model can be compared with the two Bloch balls representing the two reduced density matrices $\rho_A$ and $\rho_B$, respectively. The Bloch ball coordinates ($x_1$, $x_4$, $x_0$) may be obtained by compressing a unit sphere, our base sphere, along the $y$-axis ($x_4$-axis). The concurrence $c$ of a bipartite pure state is given by $c = \sqrt{(1 - x_0^2 - x_1^2 - x_4^2)}$, which should be the same for both Bloch balls. Hence, the two Bloch balls can carry all information about a bipartite pure state, carrying the information on both the reduced density matrices and the entanglement. The two Bloch balls can essentially complete the representation of the bipartite pure states. In comparison, the presented Bloch sphere model required the Bloch spheres to be a unit 2-sphere, on which it may be possible to represent the unitary gates in two-qubit quantum circuits as simple rotations. However, in an entangled bipartite state, only certain local gates could be represented as simple rotations, namely the $y$-rotations on the base qubit and any single-qubit gate on the fiber qubit. However, the $b$-changing local gates on the base qubit change the coordinates on both spheres (BASE(A) and ENTANGLEMENT(A)) such that this local unitary gate can not be visualized as a simple rotation on the base sphere. Our model may be viewed as the Bloch ball encoded as two unit spheres. The Bloch ball representing $\rho_A$ corresponds to the BASE(A) and ENTANGLEMENT(A) spheres, and the Bloch ball $\rho_B$ corresponds to the BASE(B) and ENTANGLEMENT(B) spheres. Specifically, the Bloch ball depicting the Bloch vector $r = (r_x, r_y, r_z) = (x_1, x_4, x_0)$ for the base qubit, where $|r| < 1$ for an entangled bipartite state, becomes the base unit sphere depicting the vector $(x_1, b, x_0)$, where $x_1^2 + b^2 + x_0^2 = 1$, and the inner entanglement sphere depicting $b^2 = x_4^2 + c^2$. That is, the $b$-coordinate of the base sphere is resolved and represented in the entanglement sphere. Hence, our Bloch sphere model explicitly shows the concurrence and its complementarity to the coherence of the base qubit (specifically, to the imaginary part $x_4$). Furthermore, each 3-sphere set {base, entanglement, fiber} can uniquely yield the bipartite state vector $|\Psi\rangle$ of Equation (1), unlike the two Bloch balls, which cannot yield the original bipartite pure state of Equation (1). For example, a bipartite pure state obtained using the state purification of the density matrix $\rho_A$ is not necessarily the same as Equation (1) since there are infinitely many such pure states [1]. Therefore, the BASE(A) and ENTANGLEMENT(A) spheres in the presented model depict the reduced density matrix $\rho_A$, the concurrence $c$, and its complementarity to the coherence, whereas the Bloch ball depicts only $\rho_A$. The strength of the Bloch balls is their simplicity.

The presented model may be simpler if, instead of using all six spheres (i.e., the two alternative sets of base, entanglement, and fiber spheres), which give two alternative descriptions of the bipartite state, only four spheres were used: BASE(A), ENTANGLEMENT(A), BASE(B), and ENTANGLEMENT(B), leaving out FIBER(A) and FIBER(B) entirely. In comparison with the two Bloch balls, these four spheres explicitly depict the concurrence $c$ and its relation to the imaginary part $x_4$ of the coherence for each base qubit and this is done at the expense of doubling the number of spheres compared to the two Bloch balls. This may be a reasonable option because the fiber spheres do not provide any measurable local information about the fiber qubit except that, with the fiber sphere, each 3-sphere set is equal to the bipartite state vector of Equation (1) and that the effect of some single-qubit unitary operations on the fiber qubit may be depicted as a simple rotation on the fiber sphere.

In the literature, different measures of quantum coherence have been proposed in the resource theory of coherence but a widely accepted and physically intuitive coherence-quantifier is still lacking yet [14,15,16]. The incoherent states are the diagonal terms of the density matrix in the reference basis (computational basis states) [14]. If $d = \sqrt{(2Tr\rho_{base}^2 - 1)}$ is used as the coherence measure [17] for the base qubit, then $d$ is complementary to the concurrence $c$ in that $d^2 + c^2 = 1$. In terms of our Bloch sphere coordinates, the coherence $d$ can be expressed as $d = \sqrt{(x_0^2 + x_1^2 + x_4^2)}$. A rising concurrence $c$ means a falling coherence $d$, and vice versa. Instead, if the off-diagonal elements of $\rho_{base}$ are used in the computational basis as the measure of coherence, which was done in this paper, then its imaginary part $x_4$ and the concurrence $c$ are related by $x_4^2 + c^2 = b^2$, which is shown in the entanglement sphere. This relation is also suggestive of the coherence–concurrence complementarity [18]. For a given $b$-coordinate of the base sphere, a rising $c$ means a falling imaginary part $x_4$, and vice versa. However, a local unitary operation on the base qubit will keep $c$ constant, while changing $x_4^2$ and $b^2$ by the same amount. A joint unitary operation on both qubits will vary all three parameters while maintaining this relation. The Bloch sphere representation could add insights into the interplay of these parameters under a joint unitary operation.

## 6. Summary

In summary, for a separable bipartite pure state, the base sphere and the fiber sphere are exactly the single-qubit Bloch spheres of each qubit. For an entangled bipartite pure state, the base sphere, together with the entanglement sphere (i.e., the $S^4$ Hopf base), represents both the local information of the base qubit (i.e, the reduced density matrix) and the non-local entanglement measure concurrence of the bipartite state. Furthermore, the fiber sphere represents the fiber qubit via a simple rotation under a single-qubit unitary operation on the qubit and completes the Bloch sphere model with the three unit spheres according to Equation (4); however, the fiber sphere seems to have no physical interpretation in an entangled bipartite state.

The two-qubit pure state was represented with two equivalent sets of three spheres, with which all local qubit information and the non-local entanglement information can be represented. Each set of Bloch spheres represent the reduced density matrix of the base qubit, as well as the entanglement property, and each qubit can be alternately assigned to the base space. The Bloch ball, corresponding to the reduced density matrix of each qubit, could be represented by two unit spheres, namely, the base and entanglement spheres, which allows displaying the coherence–concurrence complementarity on the entanglement sphere and enables depicting some local unitary gates in terms of single-qubit rotation operators. As a simplification of the presented model, the BASE(A) and ENTANGLEMENT(A) spheres may be used for qubit-A, and the BASE(B) and ENTANGLEMENT(B) spheres may be used for qubit-B, along with omitting the fiber spheres.

A python program is available in GitHub to plot the Bloch spheres given the complex amplitudes of a two-qubit pure state [12].

**References.**